\documentclass[epj]{svjour}

\usepackage{mathrsfs}
\usepackage{amsmath, txfonts}
\usepackage{graphicx,bm,booktabs,bm}
\usepackage{longtable,lscape}
\usepackage{overpic,multirow,color}
\usepackage[colorlinks,
            citecolor=blue,
            anchorcolor=green,
            menucolor=orange,
            linkcolor=red,
            filecolor=red,
            runcolor=pink,
            urlcolor=blue,
            frenchlinks=red]{hyperref}

              \allowdisplaybreaks[4]

\graphicspath{{Figures/}} %

\begin{document}
\title{Hadronic molecular states from the $K\bar{K}^*$ interaction}
\author{Pei-Liang L\"u and Jun He\thanks{Corresponding author: \emph{junhe@impcas.ac.cn}} }                    

\institute{Theoretical Physics Division, Institute of Modern Physics, Chinese Academy of Sciences,
Lanzhou 730000, China \and
Research Center for Hadron and CSR Physics,
Institute of Modern Physics of CAS and Lanzhou University, Lanzhou
730000, China}
\date{Received: date / Revised version: date}
%
\abstract{
In this work, the $K\bar{K}^*$ interaction is studied in a quasipotential Bethe-Salpeter equation approach combined with the one-boson-exchange model. With the help of the hidden-gauge
Lagrangian,  the exchanges of  pseudoscalar mesons ($\pi$ and $\eta$) and vector mesons ($\rho$, $\omega$ and $\phi$) are considered to describe the $K\bar{K}^*$ interaction. Besides the direct vector-meson exchange which can be related to the Weinberg-Tomozawa term,  pseudoscalar-meson exchanges also play important roles in the mechanism of the $K\bar{K}^*$ interaction. The poles of scattering amplitude are searched to find the molecular states produced from the $K\bar{K}^*$ interaction.  In the case of quantum number $I^G(J^{PC})=0^+(1^{++})$, a pole is found with a reasonable cutoff, which can be related to the $f_1(1285)$ in experiment. Another  bound state  with $0^-(1^{+-})$ is also produced from the $K\bar{K}^*$ interaction, which can be related to the $h_1(1380)$. In the isovector sector, the interaction is much weaker and a bound state with $1^+(1^{+})$ relevant to the $b_1(1235)$ is produced but at a larger cutoff. Our results suggest that in the hadronic  molecular state picture the $f_1(1285)$ and $b_1(1235)$ are the strange partners of the $X(3872)$ and $Z_c(3900)$, respectively.
} 
\maketitle
\section{Introduction}

After the observation of the exotic resonance structure $X(3872)$ near the $D\bar{D}^*$ threshold, more and more XYZ particles, which cannot be interpreted in the traditional quark model, are observed near the thresholds of two heavy mesons.
This suggests that there maybe exist a close relation between the XYZ particles and charmed (bottomed) meson and anticharmed (antibottomed) meson interactions, which  has attracted much attention. For example, in the charmed sector, besides the $X(3872)$, the charged exotic states $Z_c(3900)$ and $Z_b(10610)$ are often interpreted as the $D\bar{D}^*$ and the $B\bar{B}^*$ molecular states, respectively~\cite{Sun:2011uh,Sun:2012zzd,Wang:2013daa,He:2015mja}.  It is interesting to see whether it can be extended to the strange sector, $i\ e$., whether there exist the strange partners of the XYZ particles from the $K\bar{K}^*$ interaction.

There already exist some proposals of the molecular state in the strange sector. Some observed mesons, such as the $f_1(1285)$, were suggested to be from the $K\bar{K}^*$ interaction. The $f_1(1285)$ is an  axial-vector state with quantum number $I^G(J^{PC}) = 0^+(1^{++})$, mass $m = 1281.9\pm0.5$ MeV, and
width $\Gamma = 24.2\pm1.1$ MeV~\cite{Agashe:2014kda}.  In recent years, however, it has been suggested to be a dynamically generated state produced
from the $K\bar{K}^*$ interaction~\cite{Roca:2005nm,Lutz:2003fm}. Such a picture has
been extensively tested in the past decade~\cite{Aceti:2015zva,Geng:2015yta}.
For example, in Refs.~\cite{Geng:2015yta,Zhou:2014ila},  the $K\bar{K}^*$ interaction was studied in the $f_1(1285)$ channel in a chiral unitary approach. The results showed that  the $f_1(1285)$ appears as a bound state
in the dynamical picture.

To study the nature of the $f_1(1280)$ and other states near $K\bar{K}^*$ threshold, we should construct the $K\bar{K}^*$ interaction and search the molecular state from such interaction. In the chiral unitary approach a chiral invariant hidden-gauge Lagrangian is often adopted and a Weinberg-Tomozawa (WT) term is derived to describe the interaction~\cite{Roca:2005nm,Aceti:2014uea,Geng:2015yta,Zhou:2014ila}, which is identical to vector-meson exchange with some approximations~\cite{Birse:1996hd}. However, in the study of the XYZ particles the original one-boson exchange (OBE) model is widely adopted to study the molecular state~\cite{Chen:2016qju}. In the OBE model, the pion exchange is often more important than the vector-meson exchange due to  small mass of pion meson. Hence, it is interesting to study the $K\bar{K}^*$ interaction in the OBE  model and compare it with the results obtained only with the WT term.

In Ref.~\cite{He:2015mja}, the molecular state from the $D\bar{D}^*$ interaction, which is related to the $Z_c(3900)$ observed at BESIII, was studied in the OBE model and the relevant invariant mass spectrum from BESIII was also well reproduced. Both light-meson exchanges and  $J/\psi$ exchange were included in Ref.~\cite{He:2015mja}. The larger mass of  the $J/\psi$ meson ensures that the $J/\psi$ exchange potential can be reduced to a contact term by dropping out the exchange-momentum term $q^2$ in the dominator of the propagator~\cite{He:2015mja,Aceti:2014uea}. It was found that though the $J/\psi$ meson exchange is more important as suggested in the chiral unitary approach the light-meson exchanges also provide considerable contribution. In the strange sector, the WT term are  often regarded to be deduced from the vector-meson exchange~\cite{Birse:1996hd}.  However, the mass of exchanged vector mesons in the strange sector, i.e. $\rho$, $\omega$ or $\phi$ meson, is much lighter than the $J/\psi$ meson, and even comparable to the mass of  the pseudoscalar $\eta$ meson. One may wonder whether the contribution from vector-meson exchange in the strange sector is numerically close to the contribution from the WT term as in the charmed sector. on the other hand, it is interesting to study the role of the pseudoscalar-meson exchange in the $K\bar{K}^*$ interaction.

In this work, we will make an explicit study of the $K\bar{K}^*$ interaction in the OBE model including the pseudoscalar-meson exchange and the vector-meson exchange, and compare the results with those obtained from the WT term. It will be performed in a quasipotential Bethe-Salpeter equation approach, which is covariant and unitary, and has been proposed and developed to deal with the hadron-hadron interaction and the relevant hadronic molecular states, such as the $Y(4274)$, the $\Sigma_c(3250)$, and the $N(1875)$ and the LHCb pentaquarks~\cite{He:2015mja,He:2011ed,He:2012zd,He:2013oma,He:2014nxa,He:2015yva}.

This work is organized as follows. In the next section, the hidden-gauge Lagrangians are presented and adopted to derive the potential of the $K\bar{K}^*$ interaction.  In Sec 3, the scattering amplitude is obtained through solving the quasipotential Bethe-Salpeter equation with the potential obtained in section 2, and the molecular states from the $K\bar{K}^*$ interaction are explored by searching poles from the scattering amplitude. Besides, a discussion about the form factor for the exchanged meson is also given in this section.  In the last
section, a brief summary is given.

\section{\label{sec:level1} Lagrangian and $K\bar{K}^*$ interaction}

In this work, we adopt  the hidden-gauge Lagragians as in the chiral unitary approach~\cite{Roca:2005nm,Aceti:2014uea,Geng:2015yta,Zhou:2014ila}. Such formalism was well discussed and its relation to other types of effective theories was reviewed in Ref.~\cite{Birse:1996hd}. The Lagrangians for the three-meson vertices are of forms~\cite{Bando:1984ej,Bando:1987br,Nagahiro:2008cv},
\begin{eqnarray}
 \mathcal{L}_{PPV} & = &-ig~ \langle V_\mu[P,\partial^\mu P]\rangle,  \label{Eq: lagrangian1}\\
 \mathcal{L}_{VVP} & =& \frac{G}{\sqrt{2}}~\epsilon^{\mu\nu\alpha\beta}\langle\partial_\mu V_\nu \partial_\alpha V_\beta P\rangle, \label{Eq: lagrangian2}\\
 \mathcal{L}_{VVV}&=&ig ~\langle (V_\mu\partial^\nu V^\mu-\partial^\nu V_\mu V^\mu) V_\nu\rangle   ,   \label{Eq: lagrangian3}
\end{eqnarray}
where $G=\frac{3g^2}{4\pi^2f_{\pi}}$, and the coupling constant  $g=M_V/2f_\pi$ with a  $M_V\simeq 800$ MeV and $f_\pi=93$ MeV~\cite{Birse:1996hd,Aceti:2014kja}. With SU(3) symmetry, we have the octet pseudoscalar and octet vector
\begin{eqnarray}
{P}&=&\left(\begin{array}{ccc}
\frac{\pi^{0}}{\sqrt{2}}+\frac{\eta}{\sqrt{3}}+\frac{\eta'}{\sqrt{6}}&\pi^{+}&K^{+}\\
\pi^{-}&-\frac{\pi^{0}}{\sqrt{2}}+\frac{\eta}{\sqrt{3}}+\frac{\eta'}{\sqrt{6}}&
K^{0}\\
K^- &\bar{K}^{0}&-\frac{\eta}{\sqrt{3}}+\sqrt{\frac{2}{3}}\eta'
\end{array}\right),\\
{V}&=&\left(\begin{array}{ccc}
\frac{\rho^{0}}{\sqrt{2}}+\frac{\omega}{\sqrt{2}}&\rho^{+}&K^{*+}\\
\rho^{-}&-\frac{\rho^{0}}{\sqrt{2}}+\frac{\omega}{\sqrt{2}}&
K^{*0}\\
K^{*-} &\bar{K}^{*0}&\phi
\end{array}\right).\label{vector}
\end{eqnarray}

Analogously to the $D\bar{D}^*$ system, the $K\bar{K}^*$ systems under SU(3) symmetry, which are marked as $|II_z\rangle$ with $I$ and $I_z$ being the isospin and the $z$-component of the isospin, have
corresponding flavor wave functions~\cite{Sun:2011uh,Sun:2012zzd}
\begin{align}
 |1,1\rangle&= \frac{1}{\sqrt{2}}\left(|K^{*+}\bar{K}^0\rangle+c|K^+\bar{K}^{*0}\rangle\right),  \nonumber\\
 |1,-1\rangle &= -\frac{1}{\sqrt{2}}\left(|K^{*0}K^-\rangle+c|{K}^{0}K^{*-}\rangle\right),\nonumber\\
 |1,0\rangle &= -\frac{1}{2}\Big[\left(|K^{*+}{K}^-\rangle-|K^{*0}\bar{K}^0\rangle\right) +c\left(|K^{+}K^{*-}\rangle-|K^0\bar{K}^{*0}\rangle\right)\Big], \nonumber\\
 |0,0\rangle & =-\frac{1}{2}\Big[\left(|K^{*+}K^-\rangle+|K^{*0}\bar{K}^{0}\rangle\right) +c\left(|K^{+}K^{*-}\rangle+|K^0\bar{K}^{*0}\rangle\right)\Big],\label{Eq: Wavefunction}
\end{align}
where $c=\pm$ corresponds to $C$-parity $C=\mp$ respectively.
Here we use the definition $K^{(*)-}=-|1/2-1/2\rangle$.

With the above Lagrangians and wave functions, the potential for a certain state from the one-boson exchange can be obtained analogously to the $D\bar{D}^*$ interaction~\cite{Sun:2011uh,Sun:2012zzd,He:2015mja}. There are two types of diagram, the direct diagram and the cross diagram as shown in Table~\ref{flavor factor}. We would like to note that the cross diagram is not obtained from the $u$-channel contribution directly, but a deformation of the $t$ channel under  SU(3) symmetry as discussed in Ref.~\cite{He:2014nya}.  With such categorization, the potential from the one-boson exchange is written as
\begin{eqnarray}
&&\mathcal{V}_{OBE}(k_1'k_2',k_1k_2)\nonumber  \\
&=&\sum_i~{I^{d}_i}~{\cal V}^{d}_i(k_1'k_2',k_1k_2)+\sum_i~{I^{c}_i}~\mathcal{V}^{c}_i(k_1'k_2',k_1k_2),\label{Eq: VOBE}
\end{eqnarray}
where the momenta $k_{(1,2)}$ and $k'_{(1,2)}$ are  for the initial and final $K^*$($K$), respectively.
The subscript $i$ is for the different exchanged meson, including pseudoscalar mesons ${P}=\pi$ and $\eta$, and vector mesons ${V}=\rho$, $\omega$, and $\phi $. The flavor factors $I^d_i$ and $I_i^c$ for direct and cross diagrams are calculated from the Lagrangians in Eqs.~(\ref{Eq: lagrangian1}-\ref{Eq: lagrangian3})
and wave functions in Eq.~(\ref{Eq: Wavefunction}), and presented in  Table~\ref{flavor factor}.
\renewcommand\tabcolsep{0.2cm}
\renewcommand{\arraystretch}{1.6}
\begin{table}[hbtp!]
\caption{The flavor factors $I^i_d$ and $I^i_c$ for direct and cross diagrams with different exchanged mesons.
\label{flavor factor}.}
\begin{center}
	\begin{tabular}{c|ccc|ccc|ccc}\bottomrule[2pt]
&\multicolumn{3}{c|}{Direct} &\multicolumn{6}{c}{Cross}\\
&\multicolumn{3}{c|}{\scalebox{0.1}{\includegraphics[ bb=80 480 520 760,clip]{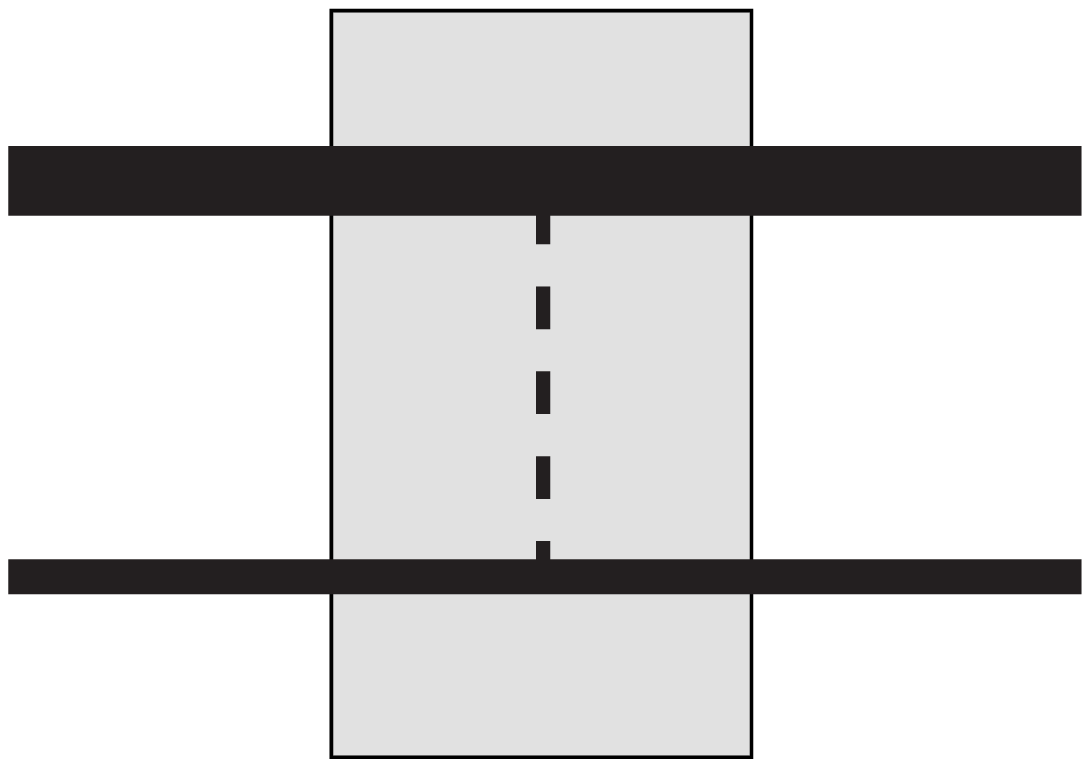}}}&\multicolumn{6}{c}{\scalebox{0.1}{\includegraphics[ bb=80 480 520 760,clip]{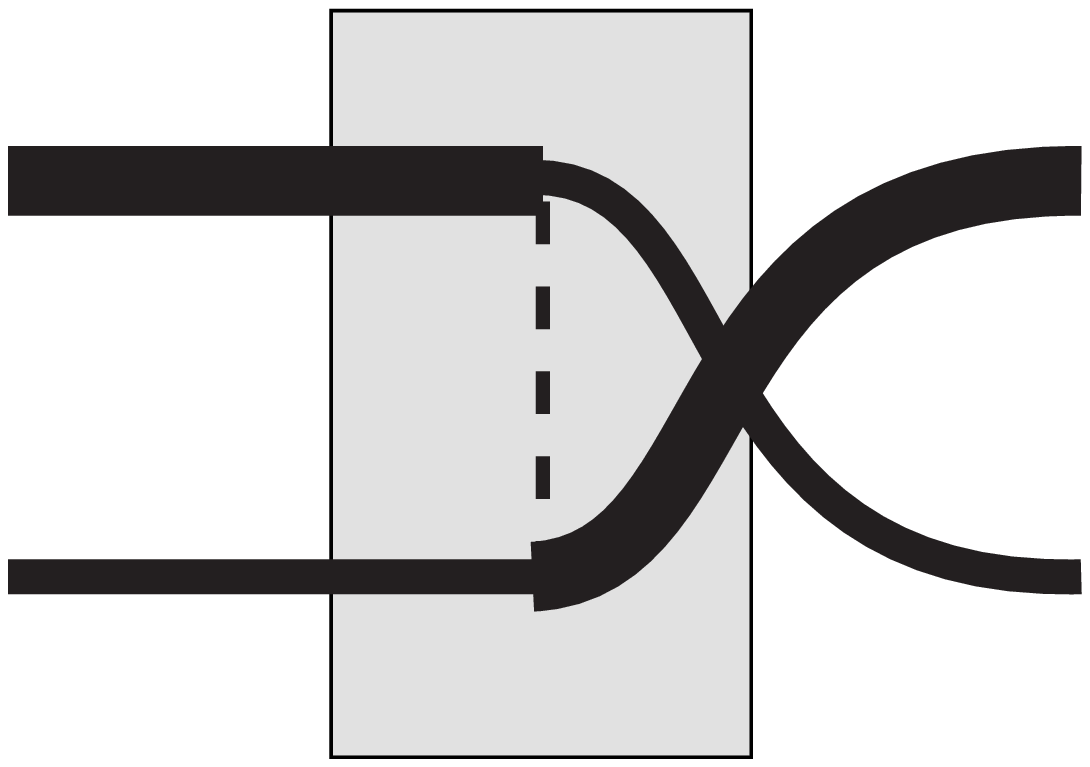}}}\\\hline
Exchanged& \multicolumn{3}{c|}{${V}$} &\multicolumn{3}{c|}{${P}$}&\multicolumn{3}{c}{${V}$}\\
meson  &  $\rho$ &$\omega$ &$\phi$ &$\pi$ &$\eta$ &$\eta'$ &  $\rho$ &$\omega$&$\phi$\\\hline
$I=1$& $-\frac{1}{2}$ & $\frac{1}{2}$ & $1$ & $-\frac{1}{2}c$ & $\frac{4}{3}c$& $\frac{1}{6}c$& $-\frac{1}{2}c$ &  $\frac{1}{2}c$ & $c$  \\
$I=0$&  $\frac{3}{2}$ & $\frac{1}{2}$ & $1$  & $\frac{3}{2}c$ &$\frac{4}{3}c$ & $\frac{1}{6}c$&$\frac{3}{2}c$ &  $\frac{1}{2}c$ &$c$\\
\toprule[2pt]
\end{tabular}
\end{center}
\end{table}

With the above preparation, the explicit forms of the ${\cal V}^{(d,c)}_i$ in Eq.~(\ref{Eq: VOBE}) read,
\begin{eqnarray}
\mathcal{V}_{{V}}^{d}&=&-{g^2}~\Big[(k'_1+k_1)\cdot\epsilon_2~(q-k_2)\cdot\epsilon'_2-(k'_1+k_1)\cdot\epsilon'_2\nonumber \\
&\cdot&(q+k'_2)\cdot\epsilon_2+(k'_1+k_1)\cdot(k'_2+k_2)~\epsilon'_2\cdot\epsilon_2\Big]~\frac{1}{(q^2-m_{{V}}^2)},\nonumber  \\
\mathcal{V}_{{P}}^{c}&=& -g^2~\epsilon'_2\cdot(k_1+q)~(q+k'_1)\cdot\epsilon_2~\frac{1}{(q^2-m_{{P}}^2)},\nonumber \\
\mathcal{V}_{{V}}^{c}&=& \frac{G^2}{2}~\Big[q^2~k'_2\cdot k_2~\epsilon'_2\cdot\epsilon_2+ k'_2\cdot q~\epsilon'_2\cdot k_2~q\cdot\epsilon_2\nonumber \\
&+&q\cdot k_2~k'_2\cdot\epsilon_2~\epsilon'_2\cdot q-q\cdot\epsilon_2~k'_2\cdot k_2~\epsilon'_2\cdot q \nonumber \\
&-&q\cdot k_2~k'_2\cdot q~\epsilon'_2\cdot\epsilon_2
-k'_2\cdot\epsilon_2~\epsilon'_2\cdot k_2~q^2\Big]~\frac{1}{(q^2-m_{{V}^2})},
\end{eqnarray}
where $m_{P}$ and $m_{V}$ are the masses of exchanged pseudoscalar and vector mesons, respectively. The momenta for exchanged meson are defined as $q=k_2'-k_2$ and $q=k_1'-k_2$ for direct and cross diagrams, respectively.

As discussed in the literature~\cite{Aceti:2014uea,Birse:1996hd,Molina:2009ct}, by dropping out the $q^2$ in the propagator, the vector-meson exchange is almost the same as the WT term, which is widely used in the chiral unitary approach. The explicit form of the WT term can be derived from the hidden-gauge Lagrangians   for the four-meson vertex~\cite{Bando:1984ej,Bando:1987br,Nagahiro:2008cv},
\begin{eqnarray}
 \mathcal{L}_{VVPP} & =& -\frac{1}{4f^2}~\langle[V^\mu,\partial^\nu V_\mu][P,\partial_\nu P]\rangle,\label{Eq: lagrangian4}
\end{eqnarray}
which gives a potential as
\begin{eqnarray}
\mathcal{V}_{WT}&=&-I_{WT}\frac{1}{4f^2}\epsilon_2\cdot\epsilon'_2(k_1+k'_1)\cdot(k_2+k'_2), \label{Eq: WT}
\end{eqnarray}
where flavor factors are $I_{WT}=-1$ and $-3$ for $I=1$ and $0$, respectively~\cite{Roca:2005nm}.  It is obviously identical to the potential from the direct $t$-channel vector-meson exchange by dropping out the $q^2$ and  neglecting the three-momenta of external vectors\cite{Molina:2009ct}. Besides,  it is interesting to see that with such approximations  other potentials from direct pseudoscalar-meson exchange and cross vector-meson exchange  vanish also. In this work, we will make a calculation with the original potential to check the effects of these approximations on the numerical results.

\section{\label{sec:level1} Numerical result}

In the current work a Bethe-Salpeter approach with a spectator quasipotential approximation will be adopted to search the molecular state from the $K\bar{K}^*$ interaction. In the  spectator quasipotential approximation~\cite{Gross:2008ps,Gross:2010qm}, one of the two particles is put onshell. The method was explained explicitly in the appendixes of Ref.~\cite{He:2015mja}. The molecular state  produced from the $K\bar{K}^*$ interaction corresponds to the pole of the scattering amplitude ${\cal M}$. The  quasipotential Bethe-Salpeter equation for the partial-wave amplitude with fixed spin-parity $J^P$ reads ~\cite{He:2015mja,He:2015cea},
\begin{eqnarray}
{\cal M}^{J^P}_{\lambda'\lambda}({\rm p}',{\rm p})
&=&{\cal V}^{J^P}_{\lambda',\lambda}({\rm p}',{\rm
p})+\sum_{\lambda''}\int\frac{{\rm
p}''^2d{\rm p}''}{(2\pi)^3}\nonumber\\
&\cdot&
{\cal V}^{J^P}_{\lambda'\lambda''}({\rm p}',{\rm p}'')
G_0({\rm p}''){\cal M}^{J^P}_{\lambda''\lambda}({\rm p}'',{\rm
p}),\quad\quad \label{Eq: BS_PWA}
\end{eqnarray}
where the sum extends only over nonnegative helicity $\lambda''$, and
the partial wave potential with  the fixed spin-parity $J^P$ can be calculated from the potential kernel ${\cal V}_{\lambda'\lambda}$ obtained in  the previous section as
\begin{eqnarray}
{\cal V}_{\lambda'\lambda}^{J^P}({\rm p}',{\rm p})
&=&2\pi\int d\cos\theta
~[d^{J}_{\lambda\lambda'}(\theta)
{\cal V}_{\lambda'\lambda}({\bm p}',{\bm p})\nonumber\\
&+&\eta d^{J}_{-\lambda\lambda'}(\theta)
{\cal V}_{\lambda'-\lambda}({\bm p}',{\bm p})],
\end{eqnarray}
where we choose the initial and final relative momenta as ${\bm p}=(0,0,{\rm p})$  and ${\bm p}'=({\rm p}'\sin\theta,0,{\rm p}'\cos\theta)$ with a definition ${\rm p}^{(')}=|{\bm p}^{(')}|$, and the $d^J_{\lambda\lambda'}(\theta)$ is the Wigner d-matrix.

 In this work we will adopt an  exponential
regularization by introducing a form factor in the propagator as
\begin{eqnarray}
	G_0({\rm p})\to G_0({\rm p})\left[e^{-(k_1^2-m_1^2)^2/\Lambda^4}\right]^2,\label{Eq: FFG}
\end{eqnarray}
with $k_1$ and $m_1$ being the momentum and mass of the charmed meson. The interested reader is referred to Ref.~\cite{He:2015mja} for further information about the regularization.
It is easy to see that no singularity will appear in the direct diagram. In the propagator of the meson exchange for the cross diagram a replacement $q^2\rightarrow-|q^2|$ needs to be made to remove the spurious singularity from the alternation of two final particles as done in Refs.~\cite{Gross:2008ps,Gross:2010qm}. In the OBE model, a form factor is usually introduced to compensate the off-shell effect of the exchanged meson as $f(q^2)=\Lambda^2_{OBE}/(\Lambda^2_{OBE}-q^2)$ with $\Lambda_{OBE}$ being the cutoff.  We will make an explicit discussion about this form factor later.

The integral equation will be solved by transforming the integral equation through discretizing the momenta $\text{p}$, $\text{p}'$ and $\text{p}''$ with the Gauss quadrature. In the current work, the coupling constants in the Lagrangians in Eqs.~(\ref{Eq: lagrangian1}-\ref{Eq: lagrangian4}) are determined, so the free parameters are two cutoffs introduced in the form factors.  The cutoff $\Lambda$ for the exponential regularization  is taken as a free parameter, and we expect it not far from 1 GeV, here in a range from 0.6 to 5 GeV. The cutoff for the exchanged meson $\Lambda_{OBE}$ is varied in a range from about 0.8 to infinity GeV. We will investigate all quantum numbers with $J=0$ and 1.

\begin{figure}[h!]
\setlength{\abovecaptionskip}{-5cm}
\includegraphics[bb=0 20 640 660,clip,scale=0.39]{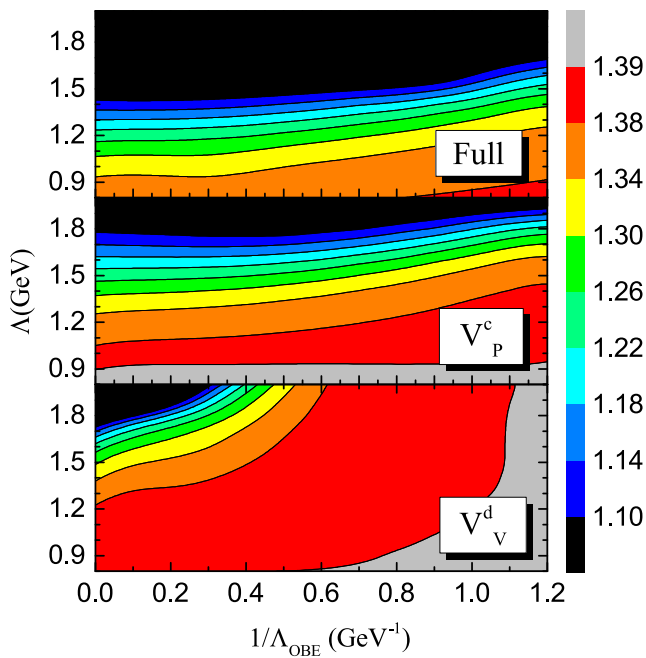}
\caption{ The pole position of the scattering amplitude for quantum number $I^G(J^{PC})=0^-(1^{++})$ with the variation of two cutoffs $\Lambda$ and $\Lambda_{OBE}$. The color pattern is for system energy $W$ of the pole in units of GeV. The upper, middle, and lower patterns are for the results of the full model, of the potential $V^c_P$ from cross pseudoscalar-meson exchanges, and of the potential $V^d_V$ from the direct vector-meson exchanges, respectively. }
\label{Fig: cutoff}
\end{figure}

Before giving the results for all quantum numbers, we present the results for the case of quantum number $I^G(J^{PC})=0^-(1^{++})$ with the variation of two cutoffs $\Lambda$ and $\Lambda_{OBE}$ in Fig.~\ref{Fig: cutoff}, which will be helpful to understand the form factor for the exchanged meson.

In the full model a bound state can be found at a regularization cutoff $\Lambda$ of about 1 GeV, and the binding energy becomes larger with the increase of the  cutoff $\Lambda$.
It is interesting to observe from Fig.~\ref{Fig: cutoff} that the result is not very sensitive to the cutoff $\Lambda_{OBE}$ for the exchanged meson. Such insensitivity can also be found in the results with only pseudoscalar-meson exchanges. Differently  from the pseudoscalar-meson exchanges, the direct vector-meson exchanges give results which change rapidly with the variation of the $\Lambda_{OBE}$ especially at small cutoffs. Such different behaviours can be related to  different masses of the exchanged mesons. The pseudoscalar  $\pi$ and $\eta$ mesons have masses much smaller than the cutoff $\Lambda_{OBE}$ which is usually chosen larger than 0.8 GeV. The potential at higher momentum transfer $q^2$ has been suppressed by the propagator of the exchanged pseudoscalar meson $1/(q^2-m_P^2)$, which makes the suppression of the form factor $\Lambda_{OBE}^2/(\Lambda^2_{OBE}-q^2)$  with a cutoff $\Lambda_{OBE}\approx 1$ GeV not so effective. On the other hand, the mass of the vector meson is about 1 GeV, which is close to the conventional value of the $\Lambda_{OBE}$. Hence,  the introduction of a form factor with a cutoff about 1 GeV will severely suppress the contribution from the vector-meson exchanges while  the pseudoscalar $\pi$ and $\eta$ meson exchanges are immune from such suppression due to their small masses.  Such conclusion can be confirmed by the insensitivity at extreme larger cutoff $1/\Lambda_{OBE}\to0$ GeV$^{-1}$ in all cases where the masses of all exchanged meson are very small compared with the cutoff.

The above analysis of results suggest that in the OBE model the vector-meson exchanges will be severely suppressed due to the introduction of the form factor with a cutoff about 1 GeV besides the suppression of the propagator itself.  It will lead to a conflict with other approaches with the WT term instead of the vector-meson exchange. In those approaches, the WT term is dominant in the interaction mechanism~\cite{Roca:2005nm,Aceti:2014uea,Geng:2015yta,Zhou:2014ila}. Now that the suppression of the vector-meson exchanges in the OBE model is from the additional form factor which is absent in the WT term, the introduction of such form factor should be reconsidered.  In the OBE model, the form factor for the exchanged meson is introduced to include the off-shellness of the exchanged meson and make the integration convergence. In our approach, the convergence is satisfied by the exponential regularization, so we need not include a form factor for this consideration. The offshellness of the exchanged meson means that the  coupling of the vector meson and the  $K^*$ should have a form factor varied with the momentum transfer $q^2$. In fact such form factor was discussed explicitly in the literature~\cite{Gasser:1984gg,Ecker:1989yg}. It was found that the pion form factor can be reproduced without additional form factor for the $\rho\pi\pi$ coupling. Instead, it is from the propagator of the vector meson. In other words, when we consider the propagator of the exchanged  vector meson, the offshellness has been included. From the above analysis, the introduction of an additional form factor for the exchanged meson in the $K\bar{K}^*$ is in fact spurious and inconsistent with chiral dynamics.

If the form factor is removed, i. e., $1/\Lambda_{OBE}\to0$ GeV$^{-1}$, the direct vector-meson exchanges become important as shown in Fig.~\ref{Fig: cutoff}. The contribution from the cross vector-meson exchanges is relatively small compared with other contributions, so we do not give their explicit results in  Fig.~\ref{Fig: cutoff}. This is because  the potential from the cross vector-meson exchange is at an order of $p/m$. Though the potentials of  pseudoscalar $\pi$ and $\eta$ exchanges are also at the same order, the small  $\pi$ and $\eta$ masses compared with vector mesons will compensate such suppression.  Comparing the results in the full model and those with cross pseudoscalar-meson exchanges and vector-meson exchanges in Fig.~\ref{Fig: cutoff}, we can found that both of the latter two contributions are important in the $K\bar{K}^*$ interaction.

As shown in Fig.~\ref{Fig: cutoff}, a bound state with a mass of about 1.285 GeV is produced from the $K\bar{K}^*$ interaction at cutoff of about 1.1 GeV, which can be related to the experimentally observed $f_1(1285)$.
In Table \ref{Tab: bound state}, we  list our results for all quantum numbers with $J=0$ and 1. The results for the pseudoscalar-meson exchanges and direct vector-meson exchanges are also presented in the same table to show their importance in the $K\bar{K}^*$ interaction. We would like to remind that the potential from the cross diagram cannot be transformed to a potential in coordinate space of a form $V(\bm r)$~\cite{He:2014nya}.  In this work, we do not consider the coupled channels, such as the $\rho\pi$ channel, which were included and  found important in other cases except the $f_1(1285)$~\cite{Roca:2005nm}.

\renewcommand\tabcolsep{0.094cm}
\renewcommand{\arraystretch}{1.46}
\begin{table}[h!]
\begin{center}
\caption{The pole position $W$ at different cutoffs $\Lambda$. The second and third columns are for the full model, the fourth and fifth columns for the cross pseudoscalar-meson exchanges, the sixth and seventh columns for the direct vector-meson exchanges, the eighth and ninth columns for the direct  vector-meson exchanges after taking $q^2\to0$, and the last two columns for the WT term. The cutoff $\Lambda$ and  energy $W$ are in units of GeV. \label{Tab: bound state}
\label{diagrams}}
	\begin{tabular}{c|cccccc|cccccccc}\bottomrule[1.5pt]
& \multicolumn{2}{c}{Full}& \multicolumn{2}{c}{$V^c_P$}&\multicolumn{2}{c|}{$V^d$}	& \multicolumn{2}{c}{$V^d|_{q^2\to 0}$}& \multicolumn{2}{c}{$V^{WT}$} \\\hline
$I^G(J^{PC})$   &  $\Lambda$ &$W$ &  $\Lambda$ & $W$  &  $\Lambda$ &$W$  &  $\Lambda$ &$W$  &  $\Lambda$ &$W$  \\\hline
$0^+(1^{++})$
& 0.6  & 1.389 & 1.0  &1.387& 1.0 &1.383 &0.8 &1.380 &0.7&1.381 \\
& 0.8  & 1.369 & 1.1  &1.377& 1.1 &1.376 &0.9 &1.366 &0.8&1.365   \\
& 1.0  & 1.326 & 1.2  &1.359& 1.2 &1.368 &1.0 &1.343 &0.9 &1.336\\
& 1.2  & 1.247 & 1.4  &1.293& 1.4 &1.343 &1.1 &1.311 &1.0&1.301 \\\hline
$0^-(1^{+-})$
& 0.8  & 1.385 & 1.2 & 1.386& 1.0 &1.383 &0.8 &1.380 &0.7&1.381  \\
& 1.0  & 1.354 & 1.3 & 1.375& 1.1 &1.376 &0.9 &1.366 &0.8&1.365  \\
& 1.1  & 1.318 & 1.4 & 1.337& 1.2 &1.368 &1.0 &1.343 &0.9 &1.336 \\
& 1.2  & 1.258 & 1.5 & 1.207& 1.4 &1.343 &1.1 &1.311 &1.0&1.301   \\\hline
$1^-(1^{+})$
&$-$ &$-$ &$-$ &$-$& 4.6 &1.388 &2.5 &1.379&1.4&1.381 \\
&$-$ &$-$ &$-$ &$-$& 4.7 &1.383 &2.6 &1.359&1.5&1.358   \\
&$-$ &$-$ &$-$ &$-$& 4.8 &1.376 &2.7 &1.332&1.6&1.336\\
&$-$ &$-$ &$-$ &$-$& 4.9 &1.366 &2.8 &1.302 &1.7&1.310\\\hline
$1^+(1^{+})$
&1.8 &1.389  & 2.7 & 1.389& 4.6 &1.388 &2.5 &1.379&1.4&1.381 \\
&1.9 &1.384  & 2.8 & 1.380& 4.7 &1.383 &2.6 &1.359&1.5&1.358   \\
&2.0 &1.372  & 2.9 & 1.343& 4.8 &1.376 &2.7 &1.332&1.6&1.336  \\
&2.1 &1.334  & 3.0 & 1.254& 4.9 &1.366 &2.8 &1.302 &1.7&1.310   \\
\toprule[1.5pt]
\end{tabular}
\end{center}
\end{table}

The results with the WT term are listed in the last two columns of Table \ref{Tab: bound state}. Because we do not consider other channels, the bound state with different $G$ parities are degenerated. Generally speaking, this results consistent with those obtained in the chiral unitary approach with four poles produced from the $K\bar{K}^*$ interaction~\cite{Roca:2005nm}. However, in the isovector section, the direct vector-meson exchange is obviously smaller than the WT term. As said in the previous section, the WT term is analytically identical to the direct $t$-channel vector-meson exchange by dropping the $q^2$ and within an approximation of neglecting the three-momenta of external vectors. To find out the origin of this difference, we give the results by dropping the $q^2$ in the propagator of the exchanged vector meson in the eighth and ninth columns of  Table \ref{Tab: bound state}.  In the isoscalar sector, the bound state appears at a cutoff $\Lambda\sim 0.8$ GeV, which is close to the result with the WT term. In the isovector sector, the cutoff needed to produce a bound state decreased obviously to about 2.5 GeV. Hence, though the WT term are consistent with the direct vector-meson exchange qualitatively, dropping out the $q^2$ will exhibit its effect in some cases especially when the interaction is weak.

Besides the state related to the $f_1(1285)$, another molecular state can be produced  with $0^-(1^{+-})$, which can be related to the $h_1(1380)$. Those two states were also produced in the coupled-channel calculation in chiral unitary approach with the WT terms~\cite{Roca:2005nm}. In the isovector sector, a bound state with $1^-(1^{+})$, which can be assigned as the $b_1(1235)$ as Ref.~~\cite{Roca:2005nm}, can be produced from the $K\bar{K}^*$ interaction  but a lager cutoff is needed. It is easily understood because the flavor factors for the isovector sector are much smaller than those for the isoscalar  sector as shown in Table \ref{flavor factor}.

\section{\label{sec:level1} Summary}

In this work, the $K\bar{K}^*$ interaction is studied in a quasipotential Bethe-Salpeter equation approach combined with the OBE model. With the help of the hidden-gauge Lagrangian, the potentials are derived including the pseudoscalar-meson and vector meson exchanges. The numerical results suggest  that the introduction of a form factor for the exchanged meson will severely suppress the contribution from the direct vector-meson exchange. The analysis suggested that the convergence of the integration and the offshellness of the exchanged meson are satisfied in our approach, so such form factor is spurious. After removing the form factor, both the pseudoscalar-meson exchanges and vector meson exchanges are important in the $KK^{*}$ interaction.

With a model without the form factor for the exchange meson, a bound state  is found with a reasonable cutoff with quantum number $I^G(J^{PC})=0^+(1^{++})$, which can be related to the $f(1285)$. Another molecular bound state  with $0^-(1^{+-})$ is also produced from the $K\bar{K}^*$ interaction, which can be related to the $h_1(1380)$. In the isovector sector, the interaction is much weaker and a bound state with $1^+(1^{+})$ relevant to the $b_1(1235)$ is produced but at a larger cutoff. The three bound states produced from the $K\bar{K}^*$ interaction in the current work can be well related to the three bound states from the $D\bar{D}^*$ interaction in Ref.~\cite{He:2015mja}, which suggests that the $f_1(1285)$ and $b_1(1235)$ are the strange partners of the $X(3872)$ and $Z_c(3900)$, respectively. The partner of the $h_1(1380)$ was also predicted in Ref.~\cite{He:2015mja}, but still not found in experiment.

\begin{acknowledgement} This project is partially supported by the Major State Basic Research Development Program in China (Grant No. 2014CB845405),
the National Natural Science Foundation of China (Grants No. 11275235 and No.11675228).
\end{acknowledgement}


\end{document}